\documentclass{emulateapj}

\slugcomment{Accepted for ApJ}

\shorttitle{Gamma-ray variability in microquasars}
\shortauthors{Owocki et al.}

\begin{document}

\title{Gamma-ray variability from wind clumping in HMXBs with jets}


\author{S.P. Owocki \altaffilmark{1}, G. E. Romero\altaffilmark{2,3},
R.~H.~D.~Townsend\altaffilmark{1,4} and A. T. Araudo\altaffilmark{2,3}}


\altaffiltext{1}{Bartol Research Institute, Department of Physics \& Astronomy,
University of Delaware, Newark, DE 19716, USA.}
\altaffiltext{2}{Inst.\ Argentino de Radioastronom\'{\i}a  (CCT La Plata, CONICET), 
C.C.5, 1894 Villa Elisa,  Buenos Aires, Argentina}
\altaffiltext{3}{Facultad de Ciencias Astron\'omicas y Geof\'\i sicas, 
Universidad Nacional de La Plata, Paseo del Bosque, B1900FWA La Plata, 
Argentina}
\altaffiltext{4}{Department of Astronomy, University of Wisconsin-Madison, 
5534 Sterling Hall, 475 N Charter Street, Madison, WI 53706, USA.}

\begin{abstract}
In the subclass of high-mass X-ray binaries known as ``microquasars", 
relativistic hadrons in the jets launched by the compact object can interact with cold 
protons from the star's radiatively driven wind, 
producing pions that then quickly decay into gamma rays.  
Since the resulting gamma-ray emissivity depends on the target density, 
the detection of rapid variability in microquasars
with GLAST and the new generation of Cherenkov imaging arrays could be
used to probe the clumped structure of the stellar wind. 
We show here that the fluctuation in gamma rays can be modeled using a 
``porosity length'' formalism, usually applied to characterize clumping 
effects. 
In particular, for a porosity length defined by $h \equiv \ell/f$, i.e. as the ratio of
the characteristic size $\ell$ of clumps to their volume filling factor $f$, 
we find that the relative fluctuation in gamma-ray emission in a binary with
orbital separation $a$ scales as
$\sqrt{h/\pi a}$ in the ``thin-jet'' limit, and is reduced by a factor
$1/\sqrt{1 + \phi a/2 \ell}$ for a jet with a finite opening angle
$\phi$.
For a thin jet and quite moderate porosity length $h \approx 0.03 \, a$, 
this implies a ca. 10\% variation in the gamma-ray emission.
Moreover, the illumination of individual large clumps might result in 
isolated flares, as has been recently  observed in some massive gamma-ray binaries. 
\end{abstract}


\section{Introduction}
\label{s_intro}

One of the most exciting achievements of high-energy astronomy in
recent years has been to establish that high-mass X-ray binaries
(HMXBs) and microquasars are variable gamma-ray sources (Aharonian et
al.  2005, 2006; Albert et al.  2006, 2007).  The variability is
modulated with the orbital period, but in addition short-timescale
flares seem to be present (Albert et al.  2007, Paredes 2008).  Since
at least some of the massive gamma-ray binaries are known to have
jets, interactions of relativistic particles with the stellar wind of
the hot primary star seem unavoidable (Romero et al.  2003).  At the
same time, there are increasing reasons to think that the winds of hot
stars have a clumped structure (e.g. 
Dessart \& Owocki 2003, 2005; 
Puls et al. 2006, and references therein). 
The observational signatures of such clumping often just depend on the
overall volume filling factor, with not much sensitivity to their scale.
Here we argue that gamma-ray astronomy can provide new constraints on the
clumped  structure of stellar winds in massive binaries with jets. 
At the same time, our analysis provides a simple formalism for
understanding the rapid flares and flickering in the light curves of
these objects.
Our basic hypothesis is that the jet produced close to the compact object
in a microquasar will interact with the stellar wind, producing
gamma-rays through inelastic $pp$ interactions, and that the emerging
gamma-ray emission will present a variability that is related to  the
structure of the wind.  Thus the detection of rapid variability by
satellites like GLAST and by Cherenkov arrays like MAGIC II, HESS II,
and VERITAS can be used as a diagnosis of the structure of the wind
itself.

\section{Jet-clump interactions}
\label{jet-clump}

\subsection{The general scenario}

\begin{figure}
\includegraphics[width=0.9\hsize]{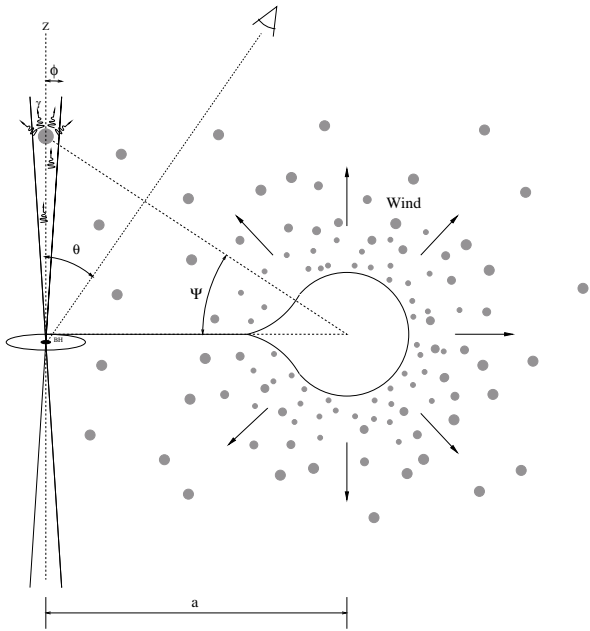} %
\caption{Sketch of the assumed model, described further in the text. }
\label{sketch}
\end{figure}

The basic scenario explored in this paper is illustrated in figure~\ref{sketch}.  
A binary system consists of a compact object
(e.g., a black hole) and a massive, hot star.  
The compact object accretes from the star and produces two jets.
For simplicity, we assume that these jets are normal to the orbital
plane and the accretion disk
(see otherwise Romero \& Orellana 2005).  
We also assume a circular orbit of radius $a$.  
The wind of the star has a clumped structure and individual clumps 
interact with the jet at different altitudes, forming an angle $\Psi$ 
with the orbital plane.
The $z$-axis is taken along the jet, forming an angle $\theta$ with the line 
of sight, with the orbit in the $xy$-plane.  
The jet has an opening angle $\phi$.  
To consider the effects of a single jet-clump interaction, we first adopt
a model for the jet\footnote{Note that the interaction of a
beam of protons and a cloud from a star has been considered before, in
the context of pulsars, by Aharonian \& Atoyan (1996).  An early
report of some material presented here can be found in Romero et al.
(2007).} (Sect. \ref{2.2}).

In addition to wind clumping, there can also be intrinsic variability 
associated with the jet.
This includes orbital modulation,
as observed in LS 5039 or LS I 61 303 
(e.g., Aharonian et al.  2006; Albert et al. 2006), and
periodic precession of steady jets (e.g., Kaufman-Bernad{\'o} et al. 2002).
Both these long-term, periodic variations would be quite distinct from 
the rapid, stochastic variations from wind clumps.
Intrinsic disturbances and shocks in jets can produce aperiodic 
variability that  might be confused with variability associated with 
jet-clump interactions. In microquasars such intrinsic fluctuations are expected 
to arise in the context of the jet-disk coupling hypothesis, 
as proposed by Falcke \& Biermann (1995) 
for the case  of AGNs, and observationally demonstrated for a galactic
microquasar by Mirabel et al. (1998).  
The same effect has been observed in AGNs (Marscher et al. 2002). 
Thus intrinsic variability in the jet would likely be preceded by  
a change in the accretion disk X-ray activity, whereas in the case of a jet-clump
interaction, the effect should be the opposite:
first the gamma-ray flare would  appear, and then, an non-thermal X-ray flare produced
by the secondary electrons and positrons as well as the primary electrons
injected into the clump would show up. 
Depending on the magnetic field and the clump density, the 
X-ray radiation could be dominated by synchrotron, inverse-Compton, or 
Bremsstrahlung emission, with a total luminosity related to that of the gamma-ray flare.
In summary,  
simultaneous X-ray observations with gamma-ray observations could be  
used to differenciate jet-clump events from intrinsic variability  
produced by the propagation of shocks in the jets.

\subsection{Basics of the jet model and jet-clump interaction}\label{2.2}

The matter content of the jets produced by microquasars is not
well-known.  
However, the presence of relativistic hadrons in the jets of SS433 has been
directly inferred from iron X-ray line observations 
(e.g. Kotani et al.  1994, 1996; Migliari et al.  2002). In addition, the large 
perturbations some jets cause in the
interstellar medium imply a significant baryon load (Gallo et
al. 2005, Heinz 2006). 
The fact that the jets are usually well-collimated also favors a
content with cold protons that provide confinement to the
relativistic gas.  
We adopt here the basic jet model proposed by Bosch-Ramon et al.  (2006), 
where the jet is dynamically dominated by cold protons.  
{\rm
Since the jet launching likely stems from
magneto-centrifugal effects (e.g., Blandford \& Payne 1982),
the jet magnetic field is assumed to be in equipartition with 
the particle energy density, with typical values of 1~kG.

Shocks from plasma collisions in the jet can produce a non-thermal
relativistic  particle population.
But only a fraction $q_{j} \approx 0.1$ of the
total jet luminosity $L_{j} \approx 10^{37} {\rm erg \, s^{-1}}$ is
expected to be converted into relativistic protons by such diffusive shock 
acceleration at the jet base
(e.g., Riegler et al.\ 2007).
The resulting gamma-ray emission can be calculated as in Romero et al. (2003)
and Orellana et al.  (2007).
For interaction between relativistic ($\sim$TeV) protons in the jet 
with cold protons in the wind, a characteristic cross section
is $\sigma \approx 3.4 \times 10^{-26}$~cm$^{2}$ 
(Kelner et al. 2006).
For a typical wind mass loss rate 
${\dot M} = 10^{-6} M_\odot {\rm yr}^{-1}$ 
and speed $v=1000 \, {\rm km \, s}^{-1}$,
the characteristic wind column depth traversed by 
the jet from an orbital separation distance
$ a \sim 0.2 AU \sim  3 \times 10^{12}~$cm
is $N \approx 5 \times 10^{22}$~cm$^{-2}$.
This implies that  only a small fraction, 
$ \tau_{w} \approx \sigma N \approx 0.002 $,
of relativistic particles in the jet are converted to gamma-rays
by interaction with the entire wind.
The leads to a mean gamma-ray luminosity of 
$L_{\gamma} = q_{j} \tau_{w} L_{j} 
\approx 2 \times 10^{33} \, {\rm erg \, s}^{-1}$.

Clumps in the wind can lead to variations and flares in this gamma-ray
emission.
For clumps of size $\ell \approx 10^{11}$~cm, corresponding to a few
percent of the stellar radius, the flow into the jet at the 
wind speed implies a flare timescale less than an hour.
While quite short, this is comparable to the variability
already detected in Cygnus X-1 by MAGIC (Albert et al. 2007). 
HESS II and MAGIC II will have a higher sensitivity, so these 
instruments should be able to detect  variability from galactic
sources like LS 5039, LS I +61 303 and Cygnus X-1 on timescales below 
an hour.

The flare brightness depends on the clump column depth and the
resulting fraction of the relativistic particle luminosity converted
to gamma rays.
For clumps of the above size with a volume filling factor $f=0.1$,
using the above wind parameters at an orbital separation distance
gives a clump column $N_{c} = 3 \times 10^{21} {\rm cm}^{-2}$.
The associated clump attenuation fraction is $\tau_{c} = \sigma N_{c} 
\approx 10^{-4}$, implying a flare gamma-ray brightness of 
$L_{\gamma} = q_{j} \tau_{c} L_{j} \approx 10^{32} \, {\rm erg \, s}^{-1}$.
Stronger flares could result when a large clump
crosses close to the base of the jet. 
Overall, if a microquasar is observed in an active state (i.e.  
when the jet is powerful), then satellite instruments like GLAST  
and ground-based Cherenkov telescopes 
should be able to detect variability down to timescales of $\sim 1$ h or
so, sufficient to measure variations associated with jet interactions 
with wind clumps.

The above picture assumes that the jet is not significantly dispersed 
or attenuated by other interactions with the wind, for example 
gyro-scattering off magnetic field fluctuations in the clumps.
Taking a characteristic temperature $T \approx 10^{4}$~K along with the above
parameters for the wind and clumps, we can estimate that at the
orbital separation distance $a=0.2$~AU, clumps have a typical thermal 
energy density 
$E_{c} = (3/2) n_{c} kT \sim 0.07 \, {\rm erg \, cm}^{-3}$, with
the corresponding equipartition magnetic field thus of order a Gauss.
For relativistic protons of Lorentz factor $\gamma$, the associated
gyroradius is just $r_{g} = 30 \gamma$~km.
Even for TeV particles with $\gamma \approx 10^{3}$, this 
is much less than the clump size, $r_{g} \ll \ell$,
implying that individual such particles should be quite
effectively deflected by such clumps.

However, this does {\em not} mean that such gyroscattering by wind clumps 
can substantially disperse the jet.
The simple reason is that the energy density of the jet completely
overwhelms that of the wind clumps.
For a jet with opening $\phi = 1^{o}$ and thus solid angle 
$\Omega = \pi \phi^{2} \approx 10^{-3}$~ster, the energy density 
at an orbital  distance $a$ is 
$E_{j} = L_{j}/(\Omega ca^{2}) \approx 4 \times 10^{4} {\rm erg \, cm}^{-3}$,
nearly a million times higher than for the clumps.
This suggests that, while clump-jet interactions may substantially
perturb or even destroy the clumps, the back-effect on the jet should 
be very small.
Moreover, while the dynamics of such clump destructions are likely to be
complex, the overall exposure of clumped wind protons to interaction 
with the relativistic protons in the jet may, to a first
approximation, remain relatively unaffected.
Overall, it thus seems reasonable to assume a simple interception model of
jet-wind interaction, withrelatively little dispersal or attenuation of 
the jet through the wind.
}

%

\section{Porosity-length scaling of gamma-ray fluctuation 
from multiple clumps}
\label{porosity}

Individual jet-clump interactions should be observable only as rare,
flaring events.  
But if the whole stellar wind is clumped, then integrated along the
jet there will be clump interactions occurring all the time,
leading to a flickering in the light curve, 
with the relative amplitude depending on the clump characteristics.
{\rm
Under the above scenario that the overall jet attenuation is small,
}
both cumulatively and by individual clumps,  
the mean gamma-ray emission should depend on the mean number 
of clumps intersected, 
while the relative fluctuation should (following
standard statistics) scale with the inverse square-root of this mean number.
But, as we now demonstrate, this mean number itself scales with the
same {\em porosity-length} parameter that has been used, for example, 
by Owocki and Cohen (2006) to characterize the effect of wind clumps
on absorption of X-ray line emission
(see also Oskinova, Hamman, and Feldmeier 2006).

Let us again consider the 
gamma-ray emission 
integrated along the jet.
{\rm Representing the relativistic particle component of the jet 
as a narrow beam}
with constant luminosity $L_{b} = q_{j} L_{j}$ along its 
length coordinate $z$,
the total 
mean
gamma-ray luminosity $L_{\gamma}$ scales 
(in the small-attenuation limit $L_{\gamma} \ll L_{b}$)
as
\begin{equation}
\left\langle  L_{\gamma} \right\rangle 
= L_{b} \sigma \int_{0}^{\infty} n(z) \, dz
    \, ,
\label{iint}
\end{equation}
where $n(z)$ is the local 
mean 
wind density
(i.e. averaged over any small-scale clumped structure),
and $\sigma$ 
{\rm is the gamma-ray conversion cross-section defined above.}

The {\em fluctuation} {\rm about} this {\em mean} emission depends 
on the properties of any wind clumps.
A simple model assumes a wind consisting entirely of clumps of
characteristic length $\ell$ and volume filling factor $f$, 
for which the mean-free-path for any ray through the clumps 
is given by the porosity length $h \equiv \ell/f$.
For a local interval along the jet $\Delta z$, the mean number of
clumps intersected is thus 
$\Delta N_{c} = \Delta z/h$, 
whereas the associated 
mean 
gamma-ray production is
given by
\begin{equation}
\left\langle 
\Delta L_{\gamma}
\right\rangle
= L_{b} \sigma n \Delta z = L_{b} \sigma n \Delta N_{c} h.
\end{equation}
But by standard statistics for finite contributions from a discrete number
$\Delta N_{c}$, the {\em variance} of this emission 
{\rm
about the mean
}
is
\begin{equation}
\left\langle \Delta L_{\gamma}^{2} \right\rangle - 
\left\langle \Delta L_{\gamma} \right\rangle^{2} 
={ L_{b}^{2} \sigma^{2} 
n^{2} 
\Delta z^{2} \over \Delta N_{c} }
= { L_{b}^{2} \sigma^{2} 
n^{2} 
h \Delta z } 
\, .
\end{equation}
Each clump-jet interaction is an independent process; thus, the
variance of an ensemble of interactions is just the sum of the
variances of the individual interactions.
The total variance is then just the integral that results from summing
these individual variances as one allows $\Delta z \rightarrow dz$.
Taking the square-root of this yields an expression for the
relative rms fluctuation of intensity,
\begin{equation}
{ \delta L_{\gamma} \over  \left < L_{\gamma} \right > }
= {\sqrt{  \int_{0}^{\infty}  n^{2} h \, dz } 
\over \int_{0}^{\infty} n \, dz
}
\, .
\end{equation}
{\rm
Note that, in this linearized analysis based on the weak-attenuation 
model for the jet, the cross-section $\sigma$ scales out of this 
fluctuation relative to the mean.
}

As a simple example, for a wind with a constant velocity and constant porosity
length $h$, the relative variation is just
\begin{equation}
{ \delta L_{\gamma} \over  \left < L_{\gamma} \right > }
= \sqrt{h/a} \, { 
\sqrt{\int_{0}^{\infty} dx/(1+x^{2})^{2}}
\over
\int_{0}^{\infty} dx/(1+x^{2})
}
= \sqrt{h/\pi a}
\, .
\end{equation}
%
%
Typically, if, say $h \approx 0.03 a$, 
then ${\delta L_{\gamma}}/{ L_{\gamma} }\approx 0.1$.
%
This implies an expected flickering at the level of 10\% for a
wind with such porosity parameters,
occuring on a timescale of an hour or less.

\section{Gamma-ray fluctuations from a finite-cone jet}
\label{finite-cone}

Let us now 
generalize this analysis to take account of
a small but {\em finite opening angle} $\phi$ for the jet cone.
The key is to consider now the {\em total number} of clumps intersecting the
jet of solid angle $\Omega \approx \phi^{2}$. 
At a given distance $z$ from the black hole origin, the cone
area is $\Omega z^{2}= (\phi z)^{2}$. 
For clumps of size $\ell$ and mean separation $L$, 
the number of clumps {\em intercepted}
by the volume $ \Omega z^{2} \Delta z$ is
\begin{equation}
\Delta N_{c} = \Delta z { \ell^{2} + \Omega z^{2} \over L^{3} } 
= { \Delta z \over h } \, \left [ 1 + (\phi z/ \ell)^{2} \right ]
\, ,
\end{equation}
where the latter equality uses the definition of the porosity length 
$h = \ell/f$ in terms of clump size $\ell$ and volume filling factor 
$f = \ell^{3}/L^{3}$.

Note that the term ``intercepted'' is chosen purposefully here, 
to be distinct from, e.g.,  ``contained''. 
As the jet area becomes small compared to the clump size, 
the average number of clumps {\em contained} in the volume would
fractionally approach zero, 
whereas the number of clumps {\em intercepted} approaches the finite, 
thin-jet value, set by the number of porosity lengths $h$ crossed in the
thickness $\Delta z$.
As such, for $\phi z \ll \ell $,
this more-general expression naturally recovers the thin-jet scaling,
$\Delta N_{c} = \Delta z/h$, used in the previous subsection.

Applying now this more-general scaling, the emission variance of this layer 
is given by
\begin{equation}
    { L_{b}^{2} \sigma^{2} n^{2} \Delta z^{2} \over \Delta N_{c} }
    = { L_{b}^{2} \sigma^{2} n^{2} h \Delta z \over 
    1 + \phi^{2} z^{2}/\ell^{2} } 
\, .
\end{equation}
Obtaining the total variance again by letting the sum become an
integral, the relative rms fluctuation of intensity
thus now has the corrected general form,
\begin{equation}
{ \delta L_{\gamma} \over  \left < L_{\gamma} \right > }
= {
\sqrt{ \int_{0}^{\infty}  n^{2} h \,  dz / ( 1 + \phi^{2} z^{2}/\ell^{2} )
}
\over 
\int_{0}^{\infty} n \, dz
}
\, .
\end{equation}


For the simple example that both the porosity length $h$ and 
clump size $\ell$ are fixed constants, 
the integral forms for the relative variation becomes
\def\bbc{{p}}
\begin{equation}
{ \delta L_{\gamma} \over  \left < L_{\gamma} \right > }
    = \sqrt{h/a} \, { \sqrt{\int_{0}^{\infty} 
    dx/[(1+ \bbc^{2} x^{2})(1+x^{2})^{2}]}
\over
\int_{0}^{\infty} dx/(1+x^{2})
}
\, ,
\end{equation}
where
$\bbc \equiv { \phi a / \ell } $
defines a ``jet-to-clump'' size parameter,
evaluated at the binary separation radius $a$.
Carrying out the integrals, we find the fluctuation from the thin-jet limit 
given above must now be corrected by a factor
\begin{equation}
C_{\bbc} = { \sqrt{1 + 2 \bbc} \over  1 + \bbc} 
\approx {1  \over \sqrt{1 +  \bbc/2 }}
\, ,
\end{equation}
where the latter simplification is accurate to within 6\% over the
full range of $\bbc$.

In the thin-jet limit $\bbc = \phi a/\ell \ll 1 $, the
correction approaches unity, as required.
But  in the {\em thick}-jet limit,
it scales as
\begin{equation}
C_{\bbc} \approx \sqrt { 2 \over \bbc } = 
\sqrt{ 2 \ell \over \phi a }  
~~~ ; ~~ \phi \gg \ell/a
\, .
\end{equation}

When combined with the above thin-jet results, the general scaling of
the fluctuation takes the approximate overall form
\begin{equation}
{ \delta L_{\gamma} \over  \left < L_{\gamma} \right > }
 \approx 
 \sqrt{ h/\pi a \over 1 + \phi a/2 \ell } 
\label{botline}
\end{equation}
wherein the numerator represents the thin-jet scaling, 
while the denominator corrects for the finite jet size.

If the jet has an opening of one degree, then
$\phi = (\pi/180) \approx 1.7 \times 10^{-2}$~radian.
If we assume a clump filling factor of say, $f = 1/10$, then
the example of the previous section for a fixed porosity length 
$h = 0.03\, a$
implies a clump size $\ell = 0.003 \, a$, and so a
moderately large jet-to-clump size ratio of $\bbc \approx 6$.
But even this gives only a quite modest reduction factor 
$C_{\bbc} \approx 0.5$, yielding now a relative
gamma-ray fluctuation of about 5\%.

The bottom line here is thus that the correction for finite cone size
seems likely to give only a modest (typically a factor two) reduction
in the previously predicted gamma-ray fluctuation levels of order 10\%.
This holds for  clump scales of
order a few thousandths of the binary separation,
and for jet cone angles of about 1 degree.
As the ratio between these two
parameters decreases (still keeping a fixed porosity length), 
the fluctuation level should decrease in proportion to the square root
of that ratio, i.e.
$
\delta L_{\gamma} \propto \sqrt{l/\phi} \propto 1/\sqrt{\bbc}
$.

\section{Conclusion}

Overall, for a given binary separation scale $a$, 
our general model for gamma-ray fluctuation due to jet interaction with 
clumped wind has just two free parameters, 
namely the porosity length ratio $h/a$, and 
the jet-to-clump size ratio $\bbc=\phi a/\ell$.
Given these parameters, then, within factors of order unity,  
the predicted relative gamma-ray fluctuation is given by eqn.~(\ref{botline}).
For reasonable clump properties with 
$h \approx 10 \ell \approx 0.03 a$, 
the fluctuation amplitude would be a few percent.

Note however that the formalism here is based on a simple model in which
all the wind mass is assumed to be contained in clumps of a single,
common scale $\ell$, with the regions between the clumps effectively
taken to be completely empty.
More realistically, the wind structure can be expected to contain clumps
with a range of length scales, superposed perhaps on the background smooth
medium that contains some nonzero fraction of the wind mass.
For such a medium, the level of gamma-ray fluctuation would likely be 
modified from that derived here, perhaps generally to a lower net
level, but further analysis and modeling will be required to quantify
this.

One potential approach might be to adopt the ``power-law porosity''
formalism developed to model the effect of such a clump
distribution on continuum driven mass loss (Owocki, Gayley, and Shaviv 2004).
This would introduce an additional dependence on the distribution
power index $\alpha_{p}$, with smaller values $\alpha_{p} \rightarrow 0$ 
tending to the smooth flow limit.
But for moderate power indices in the range $0.5 < \alpha_{p} < 1$, we can
anticipate that the above scalings should still roughly apply,
with some reduction that depends on the power index $\alpha_{p}$,
if one identifies the assumed porosity length $h$ with the strongest clumps.

Thus while there remains much further work to determine the likely
nature of wind clumping from hydrodynamical models, the basic porosity
formalism developed here does seem a promising way to characterize its broad
effect on key observational diagnostics, including the relative level 
of fluctuation in the gamma-ray emission of HMXB microquasar
systems.

\acknowledgments

We thank V. Bosch-Ramon for a critical reading of the manuscript and useful
comments. 
S.P.O.\ acknowledges partial support of NSF grant AST-0507581 and NASA
grant Chandra/TM7-8002X.
G.E.R.\ and A.T.A.\ received financial support from PICT\,13291 BID 1728/OC-AR
(ANPCyT, Argentina) and PIP 5375 (CONICET, Argentina). G.E.R. also acknowledge 
support by the Ministerio de Eduaci\'on y Ciencia 
(Spain) under grant AYA2007-68034-C03-01, FEDER funds.
R.H.D.T.\ acknowledges support of NASA grant  LTSA/NNG05GC36G.


\begin{thebibliography}{}
\bibitem[()]{682}Aharonian, F.A., \& Atoyan, A.M., 1996, Space Sci. Rev., 75, 357
\bibitem[()]{683}Aharonian, F.A., \& Atoyan, A.M., 2000, A\&A, 352, 937
\bibitem[()]{684}Aharonian, F. A., et al. (HESS Coll.), 2005, Science, 309, 746
\bibitem[()]{685}Aharonian, F. A., et al. (HESS Coll.), 2006, Science, 314, 1424
\bibitem[()]{688}Albert, J. et al. (MAGIC coll.), 2006, Science, 312, 1771
\bibitem[()]{689}Albert, J. et al. (MAGIC coll.), 2007, \apj, 665, L51
\bibitem[()]{691}Blandford, R. D. \& Payne, D. G., 1982, MNRAS, 199, 883
\bibitem[()]{692}Bosch-Ramon, V., Romero, G.E., Paredes, J.M., 2006a, A\&A, 447, 263
\bibitem[()]{693}Bosch-Ramon, V. 2007, Ap\&SS, 309, 321
\bibitem[()]{694}Dessart, L., \& Owocki, S.P., 2003, A\&A, 406, L1
\bibitem[()]{695}Dessart, L., \& Owocki, S.P., 2005, A\&A, 437, 657
\bibitem[()]{696}Falcke, H., \& Biermann, P.~L.\ 1995, A\&A, 293, 665 
\bibitem[()]{697}Gallo, E., et al., 2005 Nature, 436, 819
\bibitem[()]{698}Gregory, P.C, \& Taylor, A.R., 1978, Nature, 272, 704 
\bibitem[()]{699}Heinz, S., 2006, ApJ, 636, 316
\bibitem[()]{700}Kaufman Bernad{\'o}, M.~M., Romero, G.~E., \& Mirabel, I.~F.\ 2002,A\&A 385, L10
\bibitem[()]{701}Kelner, S.R., Aharonian, F. A., Bugayov V.V., 2006, Phys.Rev. D, 74, 034018
\bibitem[()]{703}Kotani, T., Kawai, N., Aoki, T., et al. 1994, PASJ, 46, L147
\bibitem[()]{704}Kotani, T., Kawai, N., Matsuoka, M., \& Brinkmann, W. 1996, 
PASJ, 48, 619
\bibitem[()]{706}Marscher, A.~P., Jorstad, S.~G., G{\'o}mez, J.-L., Aller, M.~F., 
Ter{\"a}sranta, H., Lister, M.~L., \& Stirling, A.~M.\ 2002, Nature, 417, 625 
\bibitem[()]{708}Migliari, S., Fender, R. \& M\'endez, M. 2002, Science, 297, 1673
\bibitem[()]{709}Mirabel, I.~F., Dhawan, V., Chaty, S., Rodriguez, L.~F., Marti, J., 
Robinson, C.~R., Swank, J., \& Geballe, T.\ 1998, A\&A, 330, L9 
\bibitem[()]{712}Orellana, M., Bordas, P., Bosch-Ramon, V., Romero, G.~E., \& 
Paredes, J.~M.\ 2007, A\&A, 476, 9 
\bibitem[()]{714}Oskinova, L., Hamman, W.-R., \& Feldmeier, A. 2006, MNRAS 372, 3130
\bibitem[()]{716}Owocki, S.P., \& Cohen, D.H., 2006, ApJ, 648, 565
\bibitem[()]{717}Owocki, S.P., \& Gayley, K.G., \& Shaviv, N. 2004, ApJ, 616, 525
\bibitem[()]{718}Paredes, J.M., 2008,  IJMP~D, 17, 1849
\bibitem[()]{722}Puls, J., Markova, N., Scuderi, S., Stanghellini, C., Taranova,
O., Burnley, A.\ W., \& Howarth, I. D. 2006, A\&A, 454, 625
\bibitem[()]{725}Rieger, F. M., Bosch-Ramon, V., \& Duffy, P. Ap\&SS, 309, 119
\bibitem[()]{726}Romero, G.E., Kaufman-Bernad{\'o}, M.M., \& Mirabel, I.F., 2002, 
A\&A, 393, L61
\bibitem[()]{728}Romero, G.E., et al., 2003, A\&A, 410, L1
\bibitem[()]{729}Romero, G.E., \& Orellana, M., 2005, A\&A, 439, 237
\bibitem[()]{732}Romero, G.E., et al., 2007, in Clumping in Hot Star Winds,
W.-R. Hamann, A. Feldmeier \& L. Oskinova, eds.
Potsdam: Univ.-Verl. URN: http//nbn-resolving.de/urn:nbn:de:kobv:517-opus-13981

\end{thebibliography}
\end{document}